\documentclass[pdflatex,sn-mathphys-num]{sn-jnl}

\usepackage{graphicx}%
\usepackage{multirow}%
\usepackage{amsmath,amssymb,amsfonts}%
\usepackage{amsthm}%
\usepackage{mathrsfs}%
\usepackage[title]{appendix}%
\usepackage{xcolor}%
\usepackage{textcomp}%
\usepackage{manyfoot}%
\usepackage{booktabs}%
\usepackage{algorithm}%
\usepackage{algorithmicx}%
\usepackage{algpseudocode}%
\usepackage{listings}%
\usepackage{array}
\usepackage{multirow}
\usepackage{lineno}
\usepackage{verbatim}
\usepackage{xcolor}

\usepackage[skip=5pt]{caption}
\usepackage{adjustbox}
\usepackage{url}
\newcommand{\ours}{Huracan}
\usepackage{svg}

\begin{document}
\title[]{\ours: A Skillful End-To-End Data-Driven System for Ensemble Data Assimilation and Weather Prediction}

\author*[1]{\fnm{Zekun} \sur{Ni}} \email{zekunni@microsoft.com}

\author[1]{\fnm{Jonathan} \sur{Weyn}}
 
\author[1]{\fnm{Hang} \sur{Zhang}}

\author[2]{\fnm{Yanfei} \sur{Xiang}}
\equalcont{Work done while at Microsoft Corporation.}

\author[1]{\fnm{Jiang} \sur{Bian}}

\author[1]{\fnm{Weixin} \sur{Jin}}
 
\author[1]{\fnm{Kit} \sur{Thambiratnam}} 

\author[1]{\fnm{Qi} \sur{Zhang}}

\author[1]{\fnm{Haiyu} \sur{Dong}}

\author[1]{\fnm{Hongyu} \sur{Sun}}

\affil[1]{\orgdiv{Microsoft Corporation}}
\affil[2]{\orgdiv{Department of Earth System Science, Ministry of Education Key Laboratory for Earth System Modeling, Institute for Global Change Studies}, \orgname{Tsinghua University}, \orgaddress{Beijing, China}}

\abstract{ 
    Over the past few years, machine learning-based data-driven weather prediction has been transforming operational weather forecasting by providing more accurate forecasts while using a mere fraction of computing power compared to traditional numerical weather prediction (NWP). However, those models still rely on initial conditions from NWP, putting an upper limit on their forecast abilities. A few end-to-end systems have since been proposed, but they have yet to match the forecast skill of state-of-the-art NWP competitors. In this work, we propose \ours{}, an observation-driven weather forecasting system which combines an ensemble data assimilation model with a forecast model to produce highly accurate forecasts relying only on observations as inputs. \ours{} is not only the first to provide ensemble initial conditions and end-to-end ensemble weather forecasts, but also the first end-to-end system to achieve an accuracy comparable with that of ECMWF ENS, the state-of-the-art NWP competitor, despite using a smaller amount of available observation data. Notably, \ours{} matches or exceeds the continuous ranked probability score of ECMWF ENS on 80.2\% of the variable and lead time combinations. Our work is a major step forward in end-to-end data-driven weather prediction and opens up opportunities for further improving and revolutionizing operational weather forecasting. 
}


\maketitle

\section{Introduction}

Operational weather forecasting has traditionally relied on a backbone of global numerical weather prediction (NWP). NWP systems consist of two components: data assimilation (DA) and the forecasting model. At each issue time, observation data newly produced after the last cycle are ingested into the data assimilation system, which combines them with the initial guesses -- forecasts from the last cycle -- to produce initial conditions for this new cycle. The forecasting model then uses physics-based numerical methods to generate forecasts based on these initial conditions. This system is not only immensely heavy in computation, requiring specialized supercomputers to run, but also inherently complex as it runs through a variety of steps involving both well-sounded physical and statistical equations as well as empirical parameterizations.

Over the past few years, machine learning- (ML) based data-driven weather prediction methods have offered a viable alternative to traditional NWP, providing more accurate forecasts while using a small fraction of computational power. This area was first explored by Weyn et al~\cite{weyn2019deep}. Within a few years of advancement, Pangu-Weather~\cite{bi2023accurate} and GraphCast~\cite{lam2023learning} became the first data-driven models to comprehensively outperform the best deterministic NWP forecasts. Shortly thereafter, ensemble forecasting models such as GenCast~\cite{price2025gencast} and AIFS-CRPS~\cite{lang2024aifscrps} showed that data-driven methods could also succeed at probabilistic forecasting.

Despite these breakthroughs, all operational ML-based models are still initialized with NWP-derived initial conditions such as those from ECMWF HRES and ENS. The data assimilation component has seen slower adoption of ML techniques. A few end-to-end data-driven DA systems have been proposed, such as Aardvark~\cite{allen2025aardvark}, FuXi Weather~\cite{sun2025fuxi} and GraphDOP~\cite{alexe2024graphdop}, but none of them has so far been able to match the forecast skill of state-of-the-art NWP competitors when coupled with a weather prediction model, highlighting the challenging nature of this problem. Overcoming these challenges could yield high rewards: Most of the forecast uncertainty comes from the initial conditions, not the forecasting model~\cite{magnusson2019dependence,harrison1999analysis}. Therefore, any progress on data assimilation may bring much bigger impact on forecast skill compared to improvements on the forecasting model. NWP DA systems are complicated by many processes including quality control, observation operators, error and covariance estimation, linear tangent physics and bias correction~\cite{bannister2017variational,ecmwf2023ifspart2}; it's reasonable to believe ML-based data-driven methods may be able to tackle these effectively.

In this paper, we present \ours{}, a skillful end-to-end data-driven system consisting of both ensemble data assimilation and ensemble weather prediction. Unlike previous works, \ours{} is the first to provide ensemble initial conditions, therefore completing the whole operational system with uncertainty estimates. At 1-degree resolution, our coupled system is able to provide ensemble weather forecasts for up to 10 days on a set of atmospheric variables including temperature, humidity, geopotential and wind at 13 pressure levels as well as surface-level temperature, wind and cloud cover. Using the widely accepted continuous ranked probability score (CRPS) metric for evaluation, forecasts provided by \ours{} are as good (within 2\% difference) or better than those from the state-of-the-art NWP system, ECMWF ENS, on 559 of the 697 (80.2\%) lead time and variable combinations, when evaluating against each model's own analysis. Even when comparing with the strongest baseline so far, which uses a hybrid approach of ECMWF ENS initial conditions and ML-based forecasts, our end-to-end forecasts are still as good or better on 65.1\% of the combinations. To the best of our knowledge, this is the first time that an end-to-end data-driven system can rival a whole NWP system while providing equally good initial conditions and forecasts in mere minutes after all observations arrive.

\section{Methods}

\subsection{Data}

The essential part of a data assimilation system is the observations. As is commonly used in NWP \cite{eyre2022assimilation} and previous ML-based DA models, we include level-1 observations from microwave and hyperspectral infrared sounders onboard the Metop, NOAA and JPSS satellites, level-2 bending angle profiles from GNSS-RO radio occultations, level-1 visible and infrared radiances from geostationary satellites as well as in-situ observations from land-based weather stations and radiosondes. According to diagnostics from NWP~\cite{dahoui2017assessing,healy2024methods}, we have included the observation types that contribute most to forecast skill, but have still omitted some other potentially impactful sources such as scatterometers or flight winds. Table \ref{tbl:input_data} summarizes the observations we are using in \ours{}. Our training set spans over years 2010 through 2022, while years 2023 and 2024 are reserved for validation and testing, respectively.

\begin{table}[htbp]
\caption{\textbf{Data sources used in \ours{}.}}
\label{tbl:input_data}
\centering
\begin{tabular}{c|c|c|c}
\toprule
\textbf{Category} & \textbf{Instrument} & \textbf{Period} & \textbf{Format/Dataset} \\
\midrule
\multirow{7}{*}{Microwave Sounders} & NOAA 15--19 AMSU-A & 2010-2024 & Level 1b~\cite{noaa_tovs} \\
& Metop A--C AMSU-A & 2010-2024 & Level 1b~\cite{eumetsat_amsua} \\
& Aqua AMSU-A & 2010-2024 & Level 1b~\cite{noaa_aqua_amsua} \\
& NOAA 15--17 AMSU-B & 2010-2014 & Level 1b~\cite{noaa_tovs} \\
& NOAA 18, 19 MHS & 2010-2024 & Level 1b~\cite{noaa_tovs} \\
& Metop A--C MHS & 2010-2024 & Level 1b~\cite{eumetsat_mhs} \\
& S-NPP, NOAA 20, 21 ATMS & 2011-2024 & Level 1b~\cite{noaa_snpp_atms,noaa_noaa20_atms,noaa_atms_sdr} \\
\midrule
\multirow{3}{*}{Hyperspectral IR Sounders} & Metop A--C IASI & 2013-2024 & Level 1c~\cite{eumetsat_iasi} \\
& S-NPP, NOAA 20, 21 CrIS & 2012-2024 & Level 1b~\cite{noaa_snpp_crisnsr,noaa_snpp_cris,noaa_noaa20_cris,noaa_noaa21_cris} \\
& Aqua AIRS & 2010-2024 & Level 1b~\cite{noaa_airs} \\
\midrule
Radio Occultation & GNSS-RO & 2010-2024 & Level 2~\cite{gnss_ro} \\
\midrule
\multirow{6}{*}{Geostationary Imagers} & GOES 16--19 ABI & 2017-2024 & Level 1b~\cite{noaa_goesr} \\
& Himawari 8, 9 AHI & 2015-2024 & Level 1b~\cite{noaa_jma_himawari} \\
& Meteosat 8--11 SEVIRI & 2010-2024 & Level 1.5~\cite{eumetsat_meteosat} \\
& Meteosat 8, 9 IODC SEVIRI & 2017-2024 & Level 1.5~\cite{eumetsat_meteosat_iodc} \\
& GOES 11--15 IMAGER & 2010-2017 & GridSat-GOES~\cite{knapp2018gridsatgoes} \\
& Global satellite composites & 2010-2017 & GridSat-B1~\cite{kenneth2011gridsatb1} \\
\midrule
\multirow{2}{*}{In-situ Observations} & Radiosondes & 2010-2024 & IGRA~\cite{durre2016igra} \\
& Land-based stations & 2010-2024 & ISD~\cite{noaa_isd} \\
\bottomrule
\end{tabular}
\end{table}

An inherent challenge associated with observation data is that, unlike analyses or initial conditions, observations are much more ``ill-behaved'' in that they can be sparse, don't necessarily align with certain grid points or time boundaries, and can have quality issues. Below we briefly outline the steps we take to process different kinds of observation data:
\begin{enumerate}
    \item Simple quality control is carried out, which includes removing data with bad quality flags or physically implausible values (brightness temperature $<100$~K or $>400$~K).
    \item Hyperspectral infrared and radio occultation data are processed into unified formats and compressed by autoencoders into 32 and 64 channels, respectively. Implementation of the autoencoders is outlined in Section \ref{autoencoder}.
    \item Level 1 satellite observations are first bilinearly interpolated to a higher-resolution intermediate grid from their native coordinates. All gridded satellite data, along with the already gridded GridSat datasets, are then conservatively interpolated to the final 1-degree resolution. Sparse in-situ observations and radio occultation embeddings are directly assigned to the closest grid point, with multiple observations being averaged at the same grid point.
    \item Satellite channels with similar nominal frequencies are merged into one, and all observations are grouped into equally-spaced time frames. Only the observations closest to the frame boundaries are used. Metadata including satellite viewing angles and time differences between the observations and frame boundaries are added as extra channels.
    \item After normalization, grid points without any available observations are assigned to zero.
\end{enumerate}

\begin{figure}[htbp]m
    \centering
    \begin{tabular}{cc}
        \includegraphics[width=6cm]{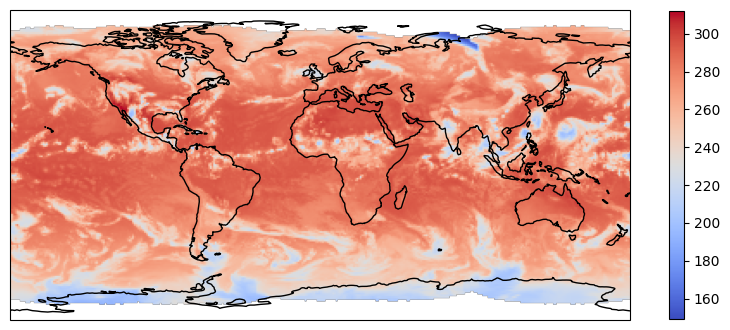} & \includegraphics[width=6cm]{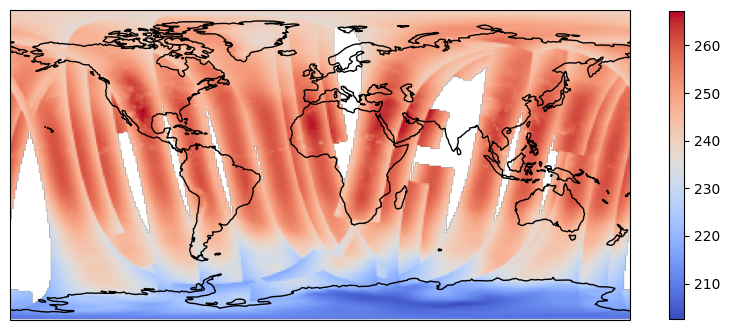} \\
        (a) & (b) \\
        \includegraphics[width=6cm]{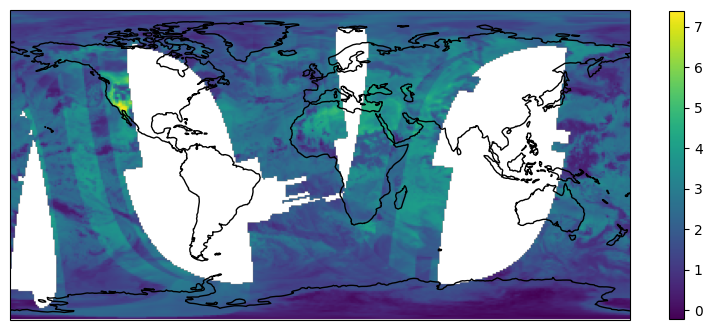} & \includegraphics[width=6cm]{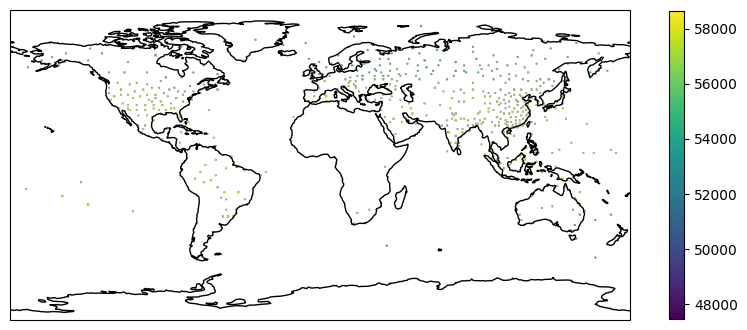} \\
        (c) & (d)
    \end{tabular}
    \caption{\textbf{Samples of observation data used in \ours{}.} (a) SEVIRI Channel 9 (ABI \& AHI Channel 14) brightness temperature; (b) ATMS Channel 6 (AMSU-A Channel 5) brightness temperature; (c) Hyperspectral IR Embedding Channel 1 between 21 UTC, August 8, 2019 and 0 UTC, August 9, 2019; (d) in-situ observation of 500 hPa geopotential at 0 UTC on August 9, 2019.}
    \label{fig:data_sample}
\end{figure}

Apart from observation data, we also utilize the ERA5~\cite{hersbach2020era5} dataset for ground truth as well as background initial conditions. \ours{} assimilates and predicts 72 variables, including 5 upper-air variables across 13 pressure levels and 7 surface variables. The full list of all model variables and input-only observation channels is provided in Table \ref{tbl:model_channels}. Samples of some observation channels are shown in Figure \ref{fig:data_sample}.

\begin{table}[htbp]
\caption{\textbf{Model variables and observation channels per time step used in \ours{}.}}
\label{tbl:model_channels}
\centering
\begin{tabular}{c|c|c}
\toprule
\textbf{Type} & \textbf{\#Channels} & \textbf{Description} \\
\midrule
\multirow{2}{*}{Model Variables} & 65 & Atmospheric variables: z, t, u, v, q on 13 levels \\
& 7 & Surface variables: 2t, 2d, 10u, 10v, tcc, tcwv, msl \\
\midrule
\multirow{2}{*}{Input-only Variables} & 1 & Insolation \\
& 2 & Constants: z, lsm \\
\midrule
\multirow{8}{*}{Observations} & $6\times11$ & SEVIRI channels 1-11, hourly\tnote{1} \\
& $2\times23$ & ATMS channels 1-22 \& MHS channel 1, 3-hourly\tnote{1} \\
& $2\times32$ & 32 hyperspectral IR embeddings, 3-hourly \\
& 64 & 64 radio occultation embeddings \\
& 65 & In-situ observations of atmospheric variables \\
& 6 & In-situ observations of surface variables except tcwv \\
& $6\times1$ & Zenith angle of the geostationary imager \\
& $2\times4$ & Zenith angles of MW and IR sounder instruments\tnote{2} \\
& $2\times4$ & Time differences of MW and IR sounder observations\tnote{2} \\
& 1 & Time difference of radio occultation profiles \\
\bottomrule
\end{tabular}
\begin{tablenotes}
    \item[1] Includes matching channels from other instruments and/or datasets.
    \item[2] One channel per time frame for each of the AMSU-A, AMSU-B/MHS, ATMS and hyperspectral IR instruments.
\end{tablenotes}
\end{table}

\subsection{Architecture}

\begin{figure}[htbp]
    \centering
\includegraphics[width=1\linewidth]{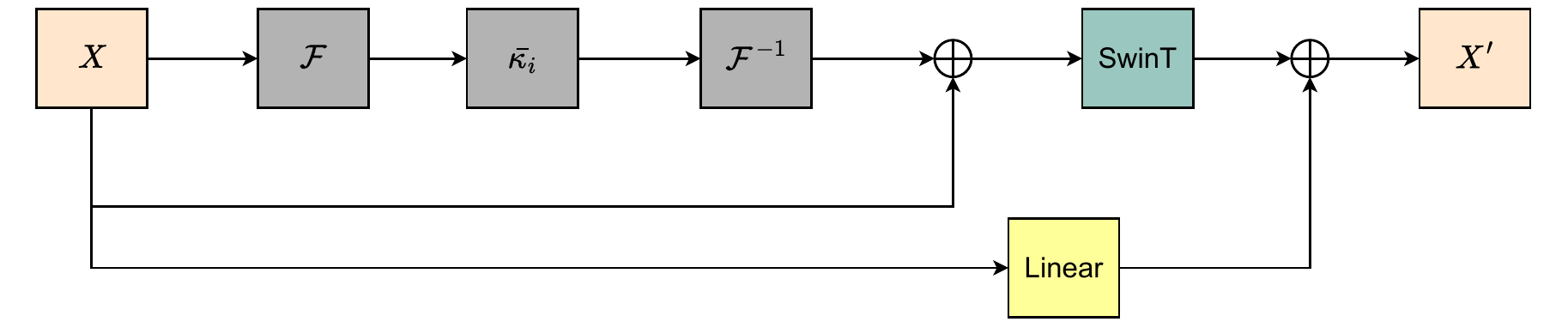}
    \caption{\textbf{The architecture of a modified SFNO block in \ours{}.} The most notable change is the replacement of a MLP block by a Swin Transformer block.}
    \label{fig:arch}
\end{figure}

\ours{} consists of two separate models for the data assimilation and forecasting tasks, but they share the identical architecture apart from the number of input channels to the encoder. The overall architecture is based on the Spherical Fourier Neural Operator (SFNO) network~\cite{bonev2023sfno}, with the following substantial modifications:
\begin{itemize}
\item In the SFNO blocks, one multi-layer perceptron (MLP) is replaced by a Swin Transformer~\cite{liu2021swin} block. This idea takes its root in the work of Kunyu~\cite{ni2023kunyu}, and has drastically improved the modeling capability of the whole network and resulted in greatly improved performance in practice. The structure of each modified SFNO block is outlined in Figure \ref{fig:arch}.
\item We have aggressively compressed the filter weights in the SFNO blocks, reducing the number of parameters 12-fold. Such compression mitigates the tendency of overfitting, and is detailed in Section \ref{sfno_compression}.
\item Similar to AIFS-CRPS~\cite{lang2024aifscrps}, in anticipation of stochasticity, noise embeddings are generated from random Gaussian noise through a MLP processor. These noise embeddings serve as the context embedding for each grid point and are passed into conditional layer normalizations~\cite{chen2021adaspeech}, replacing the existing standard layer normalizations.
\end{itemize}

Both the assimilation and forecasting models take two time steps $(T_0-6\mathrm h, T_0)$ as input and predict two time steps in the future $(T_0+6\mathrm h, T_0+12\mathrm h)$, while internally the time dimension is removed by flattening the input. Such configuration stabilizes autoregressive inference and reduces the number of autoregressive steps in rollout finetunes. Besides 72 model variables for each time step, three forcing and constant variables are added to the input, and the assimilation model additionally receives 269 channels per time step, representing all observations taking place during the past 6 hours before that time step, as shown in Table \ref{tbl:model_channels}. In this sense, the assimilation model should actually be characterized as a forecasting model with assimilation added on top.

\subsection{Training}

The whole training process consists of multiple phases. At first, a \textit{deterministic base} model is trained using 36 years of ERA5 data under L1 loss, much like every other deterministic ML-based model. This base model is then fine-tuned with the addition of noise and a change to the CRPS as the training objective. This \textit{stochastic base} model is then further fine-tuned to produce the assimilation and forecasting models:
\begin{itemize}
    \item The \textit{initial forecasting} model is produced by undergoing a 6-step autoregressive fine-tune, which means a maximum rollout window of 3 days.
    \item The \textit{initial assimilation} model is initialized with weights from the stochastic base model except for the new columns in the encoder corresponding to the observation channels, which are randomly initialized. This model is then fine-tuned using an 8-step autoregressive scheme with the initial condition and the ground truth of each step coming from ERA5 and the observation channels coming from the 15-year data mentioned above. The long rollout window of 4 days coincides with the time estimated to take for old background errors to become negligible~\cite{berre2019simulation}, and ensures the model makes the best use of the observations instead of relying heavily on the initial background from ERA5.
\end{itemize}
Finally, both models are fine-tuned together under a process consisting of 8 autoregressive assimilation steps and 8 forecasting steps. This produces the final ensemble assimilation and forecasting models which make up our \ours{} system.

Additionally, a strong baseline forecasting model is trained by fine-tuning the initial forecasting model over ECMWF analysis. This model is then inferenced with ensemble initial conditions from ECMWF ENS to produce a hybrid baseline more skillful than ECMWF ENS forecasts. We dub this system \ours{}-Hybrid.

\section{Results}

\subsection{Forecast Skill}

In this section, we select 100 issue times in the test year of 2024 from January 1 to December 17, when hindcasts from ECMWF are available, and compare CRPS scores from three ensemble forecasts over 10 lead days: ECMWF ENS, \ours{}-Hybrid and the end-to-end \ours{}, evaluated against each other's analysis.

Regarding \ours{}, we run 48 independent ensemble data assimilation (EDA) members with each member initialized from the same initial condition provided by ERA5 at 0 UTC on October 1, 2023. Even though the ultimate initial condition comes from NWP, it should have a negligible impact over our performance after at least three months of data assimilation. The ``analysis'' chosen for \ours{} is the ensemble mean of the 48 EDA members, in consistent with the practice in NWP systems. Regarding ECMWF ENS, the initial conditions, forecasts and analyses are conservatively downsampled to 1-degree to match the resolution of \ours{} and \ours{}-Hybrid.

\begin{figure}[htbp]
    \centering
\includegraphics[width=1\linewidth]{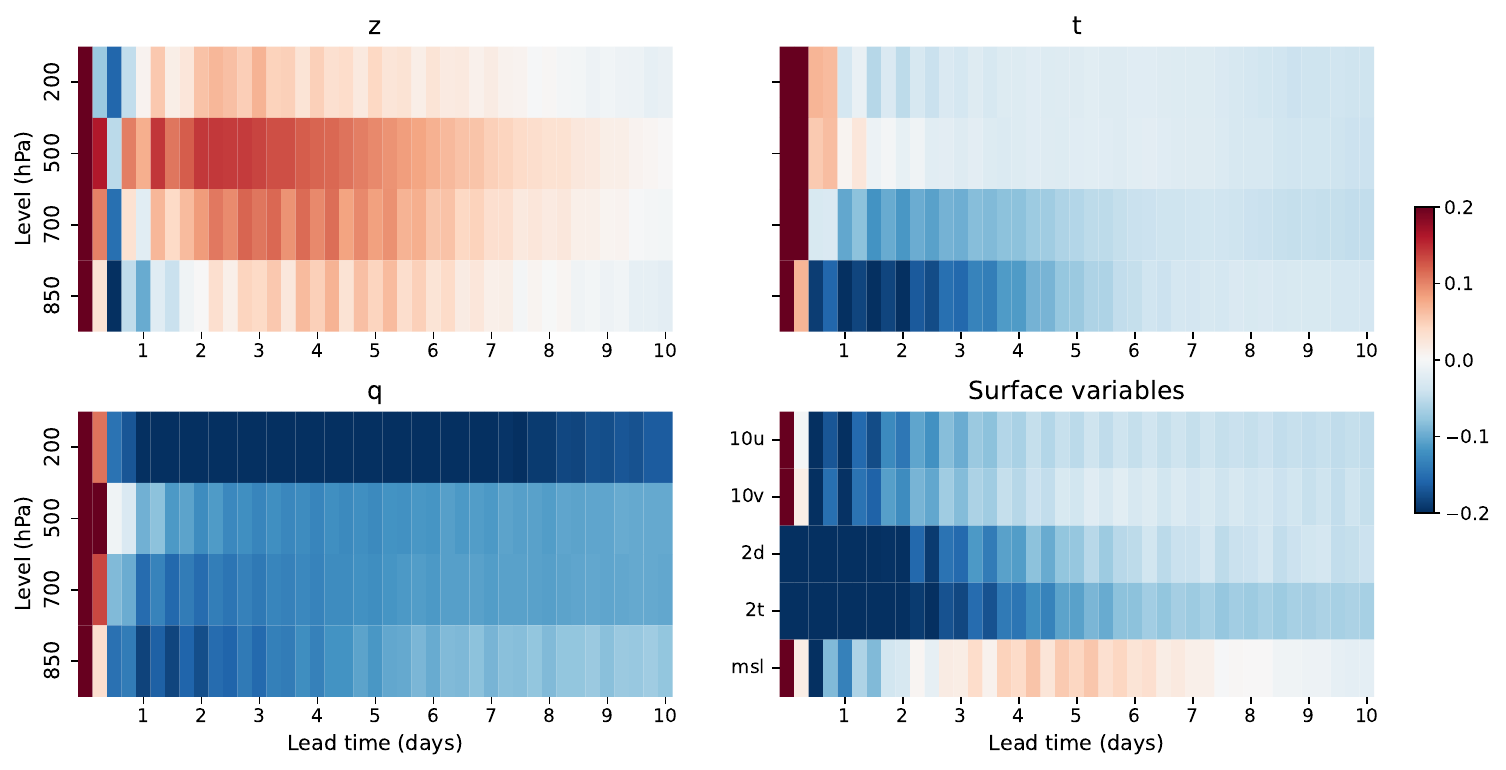}
    \caption{\textbf{Scorecard comparing CRPS scores of \ours{} versus ECMWF.} Colors are based on skill scores, which are relative differences in CRPS forecast skill. Blue colors mark improvements and red colors stand for degradations. Evaluations are carried out on geopotential (z), temperature (t) and specific humidity (q) at four vertical levels (200, 500, 700, 850 hPa) as well as 5 surface variables 2m temperature (2t), 2m dewpoint temperature (2d), 10m u and v components of wind (10u, 10v) and mean sea level pressure (msl).}
    \label{fig:scorecard}
\end{figure}

\begin{figure}[htbp]
    \centering
\includegraphics[width=1\linewidth]{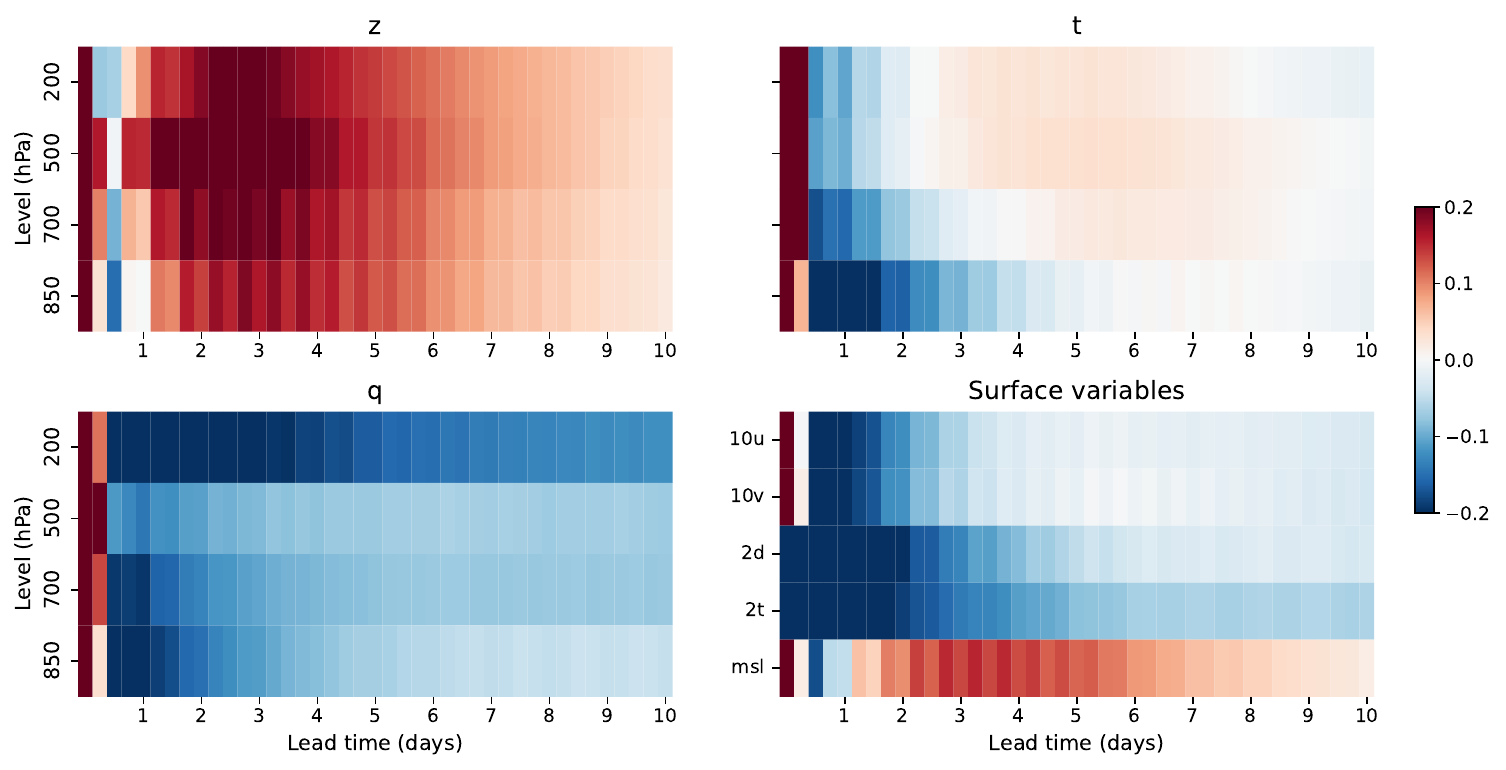}
    \caption{\textbf{Scorecard comparing CRPS scores of \ours{} versus \ours{}-Hybrid.} Like Figure \ref{fig:scorecard}, colors are based on skill scores. Blue colors mark improvements and red colors stand for degradations.}
    \label{fig:scorecard_hybrid}
\end{figure}

We adopt the notation of skill scores proposed in Graphcast~\cite{lam2023learning} and display the scorecards of CRPS skill scores in Figures \ref{fig:scorecard} and \ref{fig:scorecard_hybrid} comparing \ours{} versus ECMWF ENS, and \ours{} versus \ours{}-Hybrid.

We can conclude from the scorecards that \ours{} generally shows good forecast skill especially compared with ECMWF ENS, while there are some variations between different variables and levels. \ours{} performs better on humidity and temperature than on geopotential and pressure, and is slightly more skillful on upper, lower troposphere (200, 850 hPa) and surface compared to mid-troposphere (500 hPa). In comparison with ECMWF ENS, after removing the noisy first two steps due to initial higher spread in our ensemble, \ours{} mostly has improved metrics on temperature and humidity but is slightly behind on geopotential. However, when the baseline switches to the stronger \ours{}-Hybrid, there is more degradation in geopotential and pressure while \ours{} still generally outperforms or matches \ours{}-Hybrid on temperature, humidity and wind.

\begin{figure}[htbp]
    \centering
\includegraphics[width=1\linewidth]{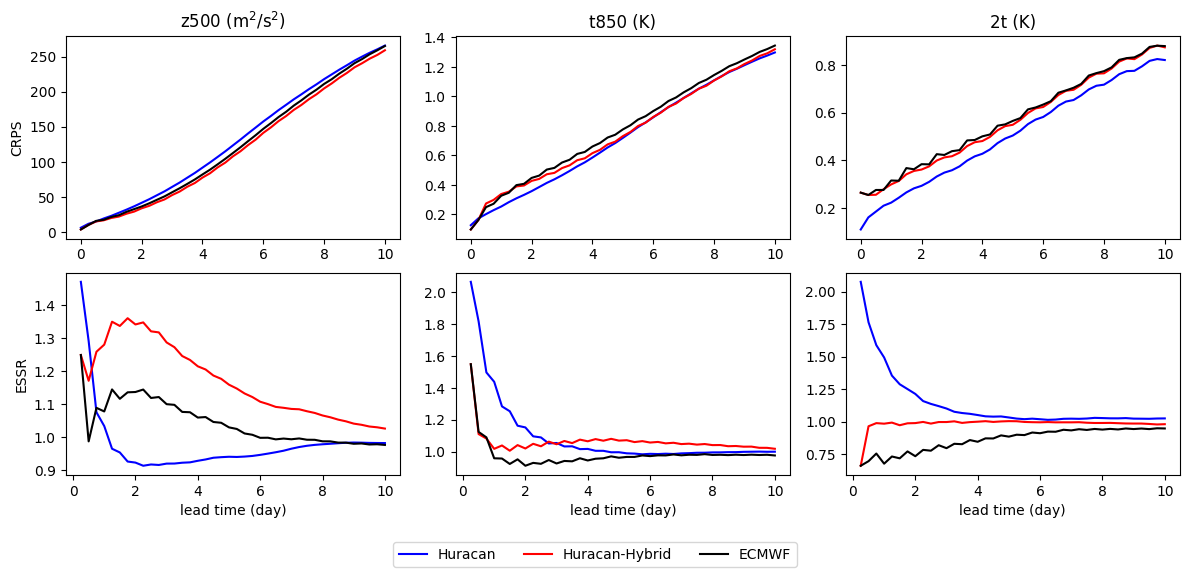}
    \caption{\textbf{Line charts detailing ensemble metrics of \ours{}, \ours{}-Hybrid and ECMWF ENS.} Charts shown above are CRPS and ESSR metrics for three key variables: 500 hPa geopotential (z500), 850 hPa temperature (t850) and 2m temperature (2t).}
    \label{fig:line_plot}
\end{figure}

To paint a more complete picture of the three ensembles, Figure \ref{fig:line_plot} shows line charts of CRPS and ensemble spread-skill ratio (ESSR) metrics of the three ensembles for three key variables. \ours{} has interesting behavior in ESSR, showing higher spread at the beginning but lower spread around 3--5 lead days. The initial higher spread should be explained by differences in methodology. in ECMWF ENS, ensemble members are generated from perturbing the deterministic control member~\cite{ecmwf2023ifspart5}, while in \ours{}, no such member exists and all members are run independently. Therefore, even though both analyses are ensemble mean, our analysis is closer to the true expectation value, therefore the larger ESSR. The likely cause of the later lower spread is overfitting, as similar evaluation results in a training year have ESSR consistently above 1 throughout the 10 lead days.

\subsection{Visualization}

\begin{figure}[htbp]
    \centering
    \begin{tabular}{c}
        \includegraphics[width=1\linewidth]{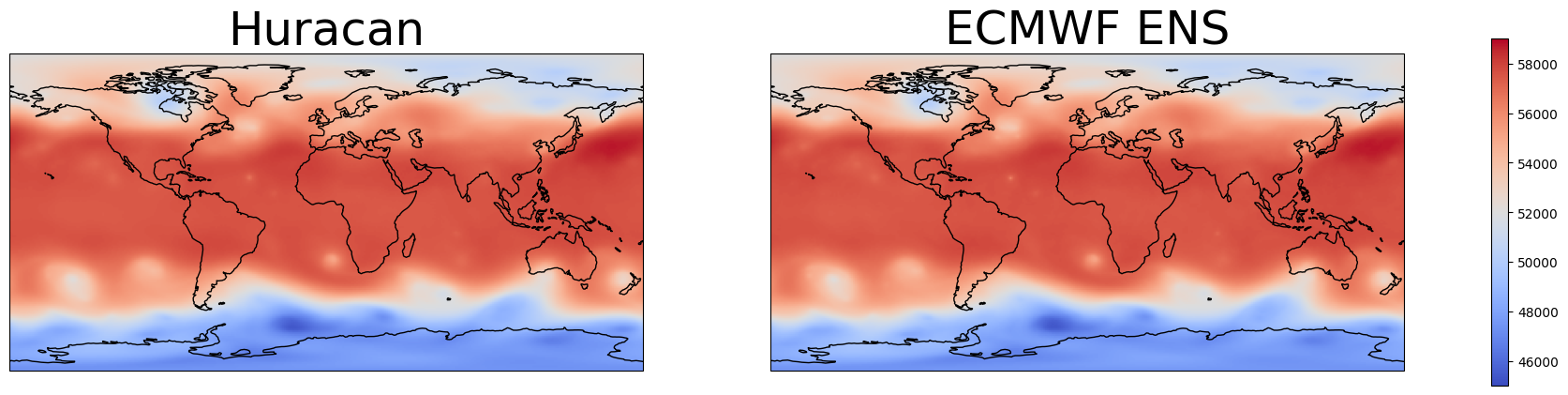}\\
        (a) Ensemble mean of 500 hPa geopotential ($\mathrm m^2/\mathrm s^2$)\\
        \includegraphics[width=1\linewidth]{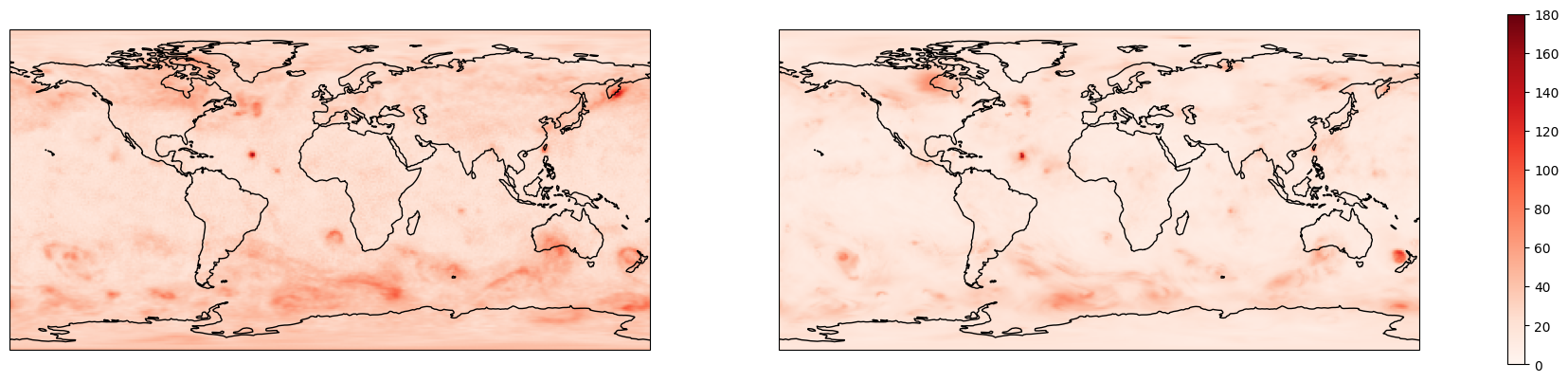}\\
        (b) Ensemble standard deviation of 500 hPa geopotential ($\mathrm m^2/\mathrm s^2$)\\
        \includegraphics[width=1\linewidth]{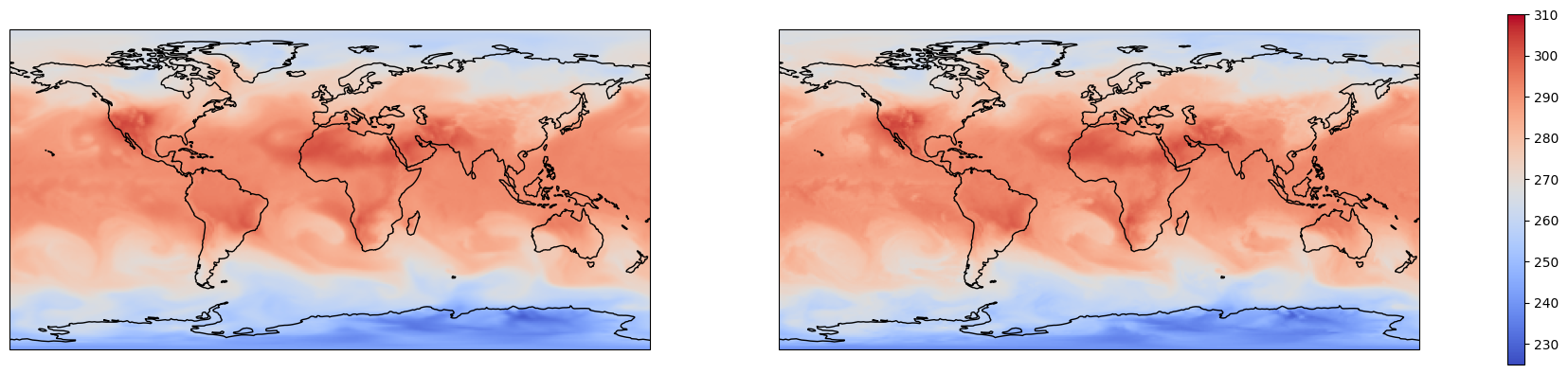}\\
        (c) Ensemble mean of 850 hPa temperature (K)\\
        \includegraphics[width=1\linewidth]{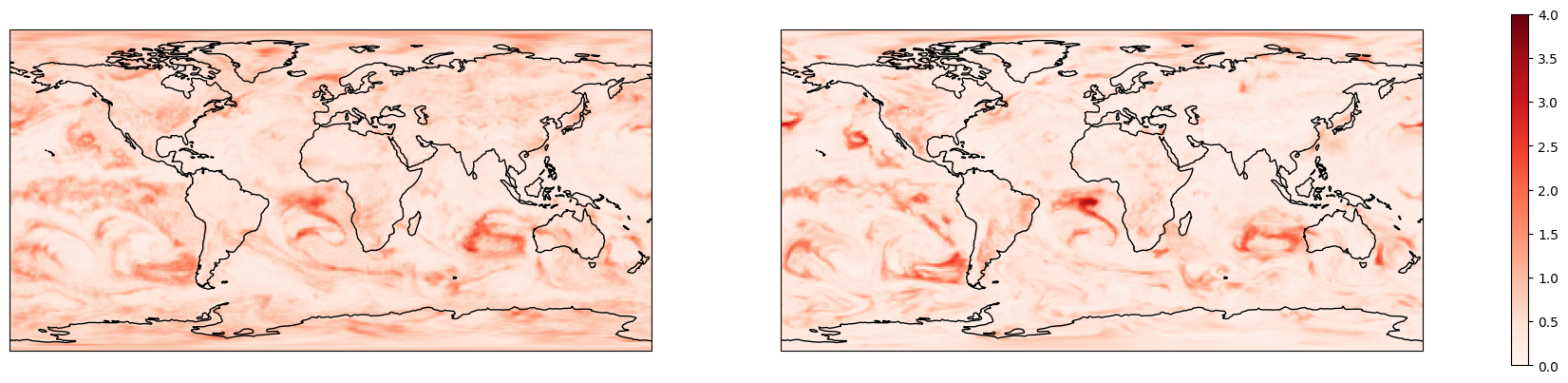}\\
        (d) Ensemble standard deviation of 850 hPa temperature (K)
    \end{tabular}
    \caption{\textbf{Visualizations of ensemble statistics.} Plots shown above are ensemble mean and standard deviation of two vital atmospheric variables, 500 hPa geopotential and 850 hPa temperature, from \ours{} and ECMWF ENS initial conditions at 0 UTC on October 3, 2024.}
    \label{fig:visualization}
\end{figure}

We additionally visualize the mean and standard deviation of \ours{} ensemble members at 0 UTC on October 3, 2024 in Figure \ref{fig:visualization}, in comparison with those from ECMWF ENS initial conditions.

The ensemble mean values generally agree well between \ours{} and ECMWF ENS, mostly showing the same large-scale patterns, with small disagreements in details. At the same time, our ensemble mean values are slightly more smoothed, due to the differences in methodology explained above. The lower native resolution of ERA5, which \ours{} is trained to predict, could also play a role, especially in the case of the Atlantic Hurricane Kirk, clearly featured in both maps.

Similar differences exist in the ensemble standard deviation values. Although both perturbations display similar patterns, our perturbations are slightly larger in general and are more spread out, while those from ECMWF ENS are more concentrated, with more extreme perturbations at a few localized spots.

\section{Discussion}

Our end-to-end ensemble data assimilation and forecasting system, \ours{}, demonstrates impressive medium-range forecast skill for both upper-air atmospheric variables and surface variables. \ours{} generally has as good or better metrics for temperature and humidity, no matter if the baseline is ECMWF ENS or the analysis-initialized \ours{}-Hybrid. Such a robust result indicates that \ours{} can effectively assimilate multiple kinds of observations, including satellite radiances and in-situ observations. Moreover, it's worth mentioning that in the case of ECMWF ENS, a 0 UTC initial condition actually comes after assimilating all observations received by 3 UTC~\cite{ecmwf2023ifspart2}; and as such, it includes future observations that \ours{} does not. The most remarkable difference though, is that our assimilation only takes one single model step, orders of magnitude faster than ECMWF ENS.

Our system relatively under-performs on pressure and geopotential. In addition to adding more data sources, there are some other, more interesting means to try improving the model skill. One potential direction is to use a better architecture to handle sparse observations, which is more in line with Aardvark and GraphDOP, rather than assigning them to grid points and zero-filling grid points without observations. A more difficult issue is that we observe big performance gaps between training set and test set. \ours{}'s performance in a training year, 2017, drastically outperforms the metrics of both \ours{} and \ours{}-Hybrid in the test year of 2024 (not shown). More research is needed to find out the cause and potential solutions to this kind of overfitting.

Another point worth discussing is the inclusion of reanalysis (ERA5) as truth data. Using ERA5 as the optimization target for both the forecasting and assimilation models means that the training of \ours{} is not independent of NWP-generated analysis, and therefore the system will learn some biases inherent in the original NWP. However, several features of our system help mitigate this, including the injection of noise and optimization of CRPS and the long auto-regressive fine-tuning which forces the model to produce more ``correct'' initial conditions that improve forecast accuracy at longer lead times. Once trained, \ours{} is nevertheless a system which is able to run completely independently of NWP indefinitely.

We believe our work of ensemble assimilation and prediction offers a promising direction towards highly skillful and operational end-to-end data-driven weather prediction systems. Our data processing and training workflow can serve as the basis of further advancements that can potentially lead to comprehensive outperformance against the current state-of-the-art approach using initial conditions from NWP.

\bibliography{paper}
\newpage
\appendix

\section{Implementation Details}

\subsection{Hyperspectral Infrared and Radio Occultation Embeddings}\label{autoencoder}

\subsubsection{Input Preparation}
\paragraph{Hyperspectral Infrared} CrIS spectrums are first apodized to match the processing level of IASI. All hyperspectral infrared spectrums are then converted to brightness temperatures and upsampled to a resolution of 0.125~cm$^{-1}$, with 16,921 channels in total.
\paragraph{Radio Occultation} Bending angle profiles from radio occultations are interpolated to 999 pressure-altitude levels from 0.001 atm to 0.999 atm. After that, six features are computed at each level, including bending angle, coordinate difference from the nearest grid point, and angle of orientation. Those features are concatenated together to yield 5,994 channels in total.

\subsubsection{Autoencoder Training}
\begin{figure}[htbp]
    \centering
    \begin{tabular}{cc}
        \includegraphics[width=6cm]{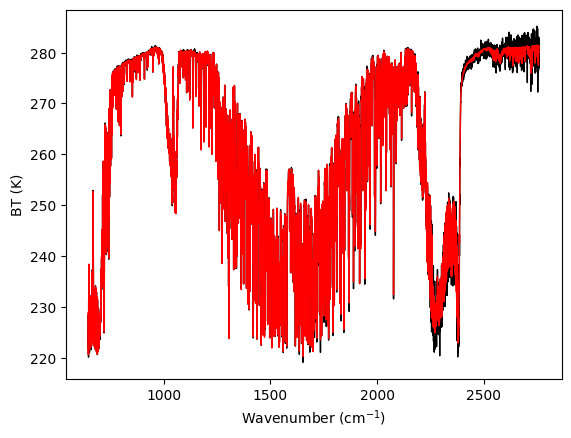} & \includegraphics[width=6cm]{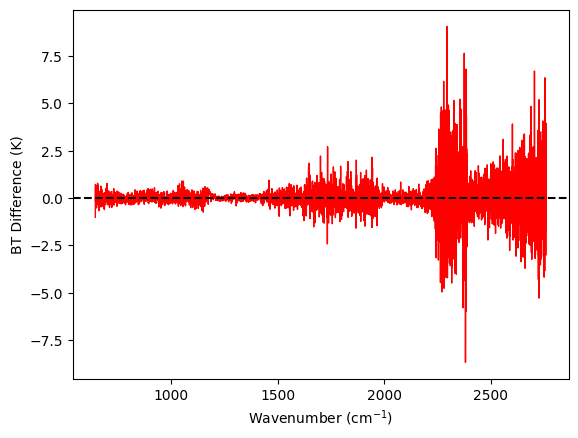} \\
        (a) & (b) \\
    \end{tabular}
    \caption{\textbf{A reconstruction example of our hyperspectral IR autoencoder.} (a) The original IASI spectrum (black) versus our reconstructed spectrum (red). The two lines mostly overlap with each other. (b) Difference of brightness temperatures (BT) between the original spectrum and our reconstruction. The reconstruction quality is poorer in the shortwave section where there is more instrument noise.}
    \label{fig:reconstruction}
\end{figure}

Two simple multi-layer perceptron (MLP) autoencoders are separately trained to compress the thousands of channels from each observation category. The two autoencoders both consist of a 4-layer encoder and a 4-layer decoder, with the size of bottleneck equal to the number of embedding dimensions (32 in hyperspectral IR, 64 in radio occultation), and are trained to minimize the L1 loss of the reconstructed spectrum. When missing channels are present, such as in the cases of data coming from instruments other than IASI, or partial occultations that do not reach the ground, those unobserved channels are set to -5 after normalization. The encoders are then used to compress hyperspectral infrared and radio occultation observations into their desired number of channels to be used in \ours{}. Figure \ref{fig:reconstruction} shows an example of hyperspectral infrared spectrum reconstruction by its autoencoder.

\subsection{SFNO Filter Compression}\label{sfno_compression}
In the original implementation of SFNO, There is one separate linear filter $\tilde\kappa_\vartheta(l)$ associated with each degree $0\le l\le L$. This brute-force implementation ignores the fact that adjacent zonal modes should be more related to each other.

In our work, we encode each degree $l$ with a sinusoidal encoding, forging connections between adjacent degrees. This encoding is then concatenated with the noise embeddings for noise injection, and transformed by a MLP into a $k$-dimensional embedding $h(l)$. The filter weight for this $l$ is in turn constructed by computing the weighted sum of learned filter weights of each embedding dimension $\tilde\kappa_i$:
$$\kappa(l)=\sum_{i=1}^kh_i(l)\tilde\kappa_i.$$
Therefore, there are only $k$ learnable linear filters instead of $L+1$ linear filters.
In our model, 1-degree resolution implies $L=179$. Setting $k=15$ results in a 12-fold reduction on the number of parameters.

\subsection{Training Parameters}
We use a batch size of 16 for all experiments and an ensemble size of 2 for training all stochastic models. Table \ref{tbl:training_setting} lists more training parameters in all training stages.

\begin{table}[htbp]
\caption{\textbf{Training settings for \ours{}.}}
\label{tbl:training_setting}
\centering
\begin{tabular}{c|c|c|c|c}
\toprule
\textbf{Training Target} & \textbf{Data} & \textbf{Max. LR} & \textbf{Schedule} & \textbf{\#Steps} \\
\midrule
Deterministic base model & 1979-2014 ERA5 & 5e-4 & cosine & 347k \\
Stochastic base model & 1979-2014 ERA5 & 2e-4 & cosine & 149k \\
Initial forecasting model & 1979-2014 ERA5 & 3e-6 & constant & 14k \\
Initial assimilation model & 2010-2022 ERA5+Obs. & 2e-4 & cosine & 46k \\
Final \ours{} models & 2010-2022 ERA5+Obs. & 3e-6 & cosine & 13k \\
\ours{}-Hybrid & 2018-2022 ECMWF HRES & 3e-6 & constant & 2.7k\\
\bottomrule
\end{tabular}
\end{table}

\end{document}